# Accelerating the timeline for climate action in California

Daniel M Kammen[1], Teenie Matlock[2], Manuel Pastor[3], David Pellow[4], Veerabhadran Ramanathan[5], Tom Steyer[6], Leah Stokes[7], and Feliz Ventura[8]

[1] Chair, Energy and Resources Group, Goldman School of Public Policy, Department of Nuclear Engineering, Director, Renewable and Appropriate Energy Laboratory, University of California, Berkeley
[2] Vice Provost for Academic Personnel, and McClatchy Chair in Communications, Department of Cognitive Science, University of California. Merced
[3] Director, USC Program for Environmental and Regional Equity (PERE), Director, USC Center for the Study of Immigrant Integration (CSII), Department of Sociology, University of Southern California
[4] Program in Environmental Studies, University of California, Santa Barbara
[5] Distinguished Professor of Atmospheric and Climate Sciences at the Scripps Institution of Oceanography, University of California, San Diego
[6] Co-President, TomKat Foundation
[7] Department of Political Science, University of California, Santa Barbara
[8] US Director, Hatch Urban Solutions

Address correspondence to kammen@berkeley.edu

## Summary
The global commitment to climate-smart actions is rising rapidly, driven by the changing political landscape, the opportunity afforded by COVID-19 recovery opportunities, and most fundamentally by the worsening climate crisis. Long an innovator in this arena, California is falling behind in its climate leadership and would benefit economically and ecologically and in terms of social justice, by establishing more aggressive goals that enable a carbon-negative economy in advance of its current 2045 target date for carbon neutrality.

## Abstract
The climate emergency increasingly threatens our communities, ecosystems, food production, health, and economy. It disproportionately impacts lower income communities, communities of color, and the elderly. Assessments since the 2018 IPCC 1.5 Celsius report show that current national and sub-national commitments and actions are insufficient. Fortunately, a suite of solutions exists now to mitigate the climate crisis if we initiate and sustain actions today. California, which has a strong set of current targets in place and is home to clean energy and high technology innovation, has fallen behind in its climate ambition compared to a number of major governments. California, a catalyst for climate action globally, can and should ramp up its leadership by aligning its climate goals with the most recent science, coordinating actions to make 2030 a point of significant accomplishment. This entails dramatically accelerating its carbon neutrality and net-negative emissions goal from 2045 to 2030, including advancing clean energy and clean transportation standards, and accelerating nature-based solutions on natural and working lands. It also means changing its current greenhouse gas reduction goals both in the percentage and the timing: cutting emissions by 80 percent (instead of 40 percent) below 1990 levels much closer to 2030 than 2050. These actions will enable California to save lives, benefit underserved and frontline communities, and save trillions of dollars. This rededication takes heed of the latest science, accelerating equitable, job-creating climate policies. While there are significant challenges to achieving these goals, California can establish policy now that will unleash innovation and channel market forces, as has happened with solar, and catalyze positive upward-scaling tipping points for accelerated global climate action.

## 1. California must lead on climate

As the September 13, 2020, front page banner headline of *The Los Angeles Times* declared: "California's climate apocalypse. Fires, heat, air pollution: The calamity is no longer in the future – it's here, now."

With a record-shattering 4.3 million acres burned (approximately 4% of the state), 32 direct deaths, and an estimated 1200 - 3000 deaths from toxic smoke that covered much of the state for several weeks,[1] California finds itself at the frontline of climate destruction. California's 2018 wildfires, less than half the size of the 2020 conflagrations, cost a staggering $148.5 billion in damages (about two thirds of California's pre-COVID 2020 state budget), with $27.7 billion (19%) in capital losses, $32.2 billion (22%) in health costs and $88.6 billion (59%) in indirect losses with a majority of those far from the actual wildfire footprint.[2] The mental health impacts of direct as well as indirect exposure to fires, such as PTSD and depression, are just coming into focus, and are potentially enormous.[3]

While touring a Sierra foothill fire zone in early September, Governor Newsom acknowledged, "We are in a climate damn emergency." He conceded that "across the entire spectrum, our climate goals are inadequate. We have to step up our game. As we lead the nation in low carbon green growth, we'll have to fast track our efforts."[4]

As of September 2020, the state has experienced a degree of wildfire activity that California's Fourth Climate Change Assessment initially forecasted to not occur until 2050.[5] This climate disaster is resulting in loss of human life and significant expense to the state, businesses, local communities, and families,[6] and is reflected globally in the record-setting heatwaves in Siberia,[7] fires and massive biodiversity loss in Australia,[8] and changes in fish populations in the Pacific Northwest.[9]

---

COVID pummeled the state budget, causing a 25% deficit in 2020 (an estimated $56 billion) and stalled the impetus for the increased investment required to catalyze speed and scale climate solutions. Yet the costs of inaction and inadequate action massively outstrip the costs of action.[10]

Years of drought, increased temperatures, and poor land and water management decisions have combined to create a tinderbox in the nation's most populous state of 40 million residents. California provides one third of the country's vegetables and two thirds of its fruits and nuts. California is a global center for technological innovation, academic excellence, and international partnerships. Each of these distinguishing features are imperiled by the climate crisis. The state is now in the early stages of a megadrought, a multi-decadal drought made severe by the growing impacts of climate change which threatens further social and economic devastation.[11]

Historically the U.S. and the world have looked to California, the world's fifth largest economy, to lead on climate-smart policies, and to continue serving as a premiere innovator and implementer of climate solutions. In 2002, for example, California's Assembly Bill 1493, requiring a reduction in GHG emissions from vehicle tailpipes, became state law. It was later adopted by 13 other states, and in 2010 became the basis for national clean car standards. When former President Trump announced the U.S. withdrawal from the Paris Climate Accord, California joined the German state of Baden-Württemberg and others to launch the Under2 Coalition in 2017. The list of signatories committing to significant GHG cuts by 2050 has grown to over 220 jurisdictions which combined encompasses over 1.3 billion people and 43% of the world economy.[12]

The clear recognition that we are not doing enough is being echoed nationally as well. The day after President Biden's inauguration, U.S. climate envoy John Kerry told an international group of business leaders that every nation has fallen behind in meeting the goals of the Paris Agreement. "No country and no continent is getting the job done," Kerry said.[13]

**California must now accelerate its climate policy innovation and implementation timelines to decarbonize the economy more rapidly.** Still, decarbonization measures, while essential, will take two to three decades to have an impact on the steeply warming curve. The need for speed is great and it is a race against time to keep warming from shooting past 2°C well before 2050.
Given the likelihood that warming will cross the 1.5°C threshold within the next 10 years,[14] and the fact that current greenhouse emissions are about 49 gigatons of CO2eq per year globally, relying just

---

on deep decarbonization is not enough anymore to limit warming well below 2°C. We need to pull on three levers to bend the warming curve before it reaches 2°C.[15]

1. The first lever, of course, is zero emissions of $CO_2$.

2. The second lever involves drastic reductions in super pollutants that are short-lived—black carbon, methane, tropospheric ozone, and hydrofluorocarbons (HFCs).[16] These super pollutants are about 30 to 2000 times more potent than $CO_2$ in trapping infrared heat. Black carbon is soot emitted by mostly diesel engines and lives in the air for a week; its heat-trapping power is 2000 times that of $CO_2$. HFCs used as refrigerants are also about 2000 time more potent. Collectively these super pollutants are responsible for about 40% of warming globally. Reducing methane emissions by half, reducing soot emissions by 80% with soot-free vehicles such as electric vehicles, replacing currently used HFCs with zero- to low-warming potential refrigerants, and decreasing sources of methane emissions such as leaks from natural gas pipes, food, and other landfilled organic waste, if implemented now, can cut the rate of warming over the next 2 to 3 decades by half.

3. The third lever involves extraction of the $CO_2$ already released into the atmosphere from human activities. Many studies show that as much as 500 to 1000 billion tons of $CO_2$ have to be extracted.

In addition to decarbonizing, California should rapidly reduce its emissions of super pollutants including methane, black carbon soot and HFCs used as refrigerants, to bend the warming curve by 2035.[17] California took a step in this direction in 2016 with SB1383 that focuses on super pollutants, but more is needed.[18] The targets set forth by SB1383 should be expedited. Accelerated reduction of GHGs and super pollutants has the potential to create hundreds of thousands of well-paying jobs while securing an equitable, clean energy future for all.[19]

The COVID-19 crisis, while devastating, has produced a pause in emissions growth that means if we act forcefully, we may have already seen peak emissions.[20] Doing so would create a healthy energy future for all and large reductions in local air pollution including ozone.[21]

---

[15] Veerabhadran Ramanathan, Juliann E. Allison, Maximilian Auffhammer, David Auston, Anthony D. Barnosky, Lifang Chiang, William D. Collins, Steven J. Davis, Fonna Forman, Susanna B. Hecht, Daniel Kammen, C.-Y. Cynthia Lin Lawell, Teenie Matlock, Daniel Press, Douglas Rotman, Scott Samuelsen, Gina Solomon, David G. Victor, Byron Washom, 2015 Executive Summary of the Report, Bending the Curve:10 scalable solutions for carbon neutrality and climate stability. Published by the University of California, October 27, 2015

[16] M. Molina, V. Ramanathan, & D. Zaelke. Best path to net zero: Cut short-lived super-pollutants. Bulletin of the Atomic Scientists. April 2, 2020

[17] Xu, Y. and V. Ramanathan, Well below 2°C:Mitigation strategies for avoiding dangerous to catastrophic climate changes. 2017; Proceedings of the National Academy of Sciences, vol. **114 (39),** 10315–10323.

[18] https://www.calrecycle.ca.gov/organics/slcp
https://leginfo.legislature.ca.gov/faces/billNavClient.xhtml?bill_id=201520160SB1383

[19] https://laborcenter.berkeley.edu/putting-california-on-the-high-road-a-jobs-and-climate-action-plan-for-2030/

[20] Daniel M. Kammen (2020) Over the hump: Have we reached the peak of carbon emissions? *Bulletin of the Atomic Scientists*, 76:5, 256-262. DOI: 10.1080/00963402.2020.1806585

[21] https://laborcenter.berkeley.edu/putting-california-on-the-high-road-a-jobs-and-climate-action-plan-for-2030/

## 2. Current climate targets are insufficient to secure a safe future

Since 2018, the goals recommended by the United Nation's (UN) Intergovernmental Panel on Climate Change (IPCC) in their 2018 special report on "Global Warming of 1.5°C" set the bar for climate ambition in countries, states, businesses, universities, and other entities.[22] As established by the UN Framework Convention on Climate Change on the heels of the 2015 historic Paris Agreement, the 1.5°C pathway report calls for the world to move beyond the initial 2°C target and limit warming to 1.5°C to avoid the worst consequences.

The 2018 IPCC prescription for remaining or returning to under the 1.5°C threshold of dangerous global warming includes cutting emissions by 45% by 2030 (from 2010 levels) and investing in carbon dioxide removal (CDR). The latter includes restoration of natural ecosystems, reforestation, and soil carbon sequestration, with their many co-benefits, to draw down anthropogenic carbon additions to the atmosphere (up to 1,000 Gt of carbon dioxide equivalents) and achieve net-zero emissions by 2050.

As bold as that goal is, it is clear that it is not enough. While some will push back on the call here to make 2030 the new target date, global warming is today approaching 1.2°C, with 2030 now projected as the timeframe when we will exceed 1.5°C.[23] The UN operates by consensus and all communications about its scientific recommendations are filtered through political decision makers from member nations, resulting in inherently conservative policy guidance. The need to accelerate this process means that states and nations with the ability to act must demonstrate through their actions that faster progress is possible, affordable, and equitable.

Even with the economic slowdown resulting from the coronavirus pandemic, emissions are projected to drop temporarily by only about 7%.[24] This drop translates into a 0.01°C reduction in global warming by 2050, according to the United Nations Environment Programme's (UNEP) 2020 Emissions Gap Report. Per UNEP, the world must triple its emissions reductions commitments to follow a 2°C (3.6°F) pathway and multiply them *five* times to follow a 1.5°C (2.7°F) pathway to achieve the goal of limiting global warming to the 2015 Paris Climate Agreement targets.[25] Emissions of $CO_2$ and super pollutants must fall by a staggering 7.6% each year between now and 2030 to stay on track with 1.5°C of warming, according to the report.

---

New scientific findings show a rapidly worsening climate with nine of fifteen global tipping points already approaching or beyond the natural limits.[26] Said the study's co-author, "We have underestimated the risks of unleashing irreversible changes where the planet self-amplifies global warming. Scientifically, this provides strong evidence for declaring a state of planetary emergency, to unleash world action that accelerates the path towards a world that can continue evolving on a stable planet."[27]

Some recent studies shortened the timeframe for mitigating dangerous climate change (1.5°C) to between 2027 and 2042,[28] which is one to two decades earlier than the timeframe in the IPCC's 1.5°C report. Another team concluded that to secure "a tolerable climate future," we must immediately and aggressively pursue *carbon neutral energy production by 2030*. They further concluded that we must "hope for some luck" that the global climate system's sensitivity to the continued addition of warming greenhouse gas emissions is low. These conclusions were based on an assessment of more than 5 million future climate pathways.[29] A more recent study found that surface air temperatures are already around 1.25°C above the 1850-1900 pre-industrial period.[30]

Time is running out. The likelihood of abrupt, irreversible, runaway climate chaos increases every day that emissions are not reduced significantly and fast. To ensure climate, health, and economic stability, California and governments worldwide at every level must commit to and follow an extremely accelerated timeframe for GHG emissions including nature-based sequestration and ecosystem protection.

---

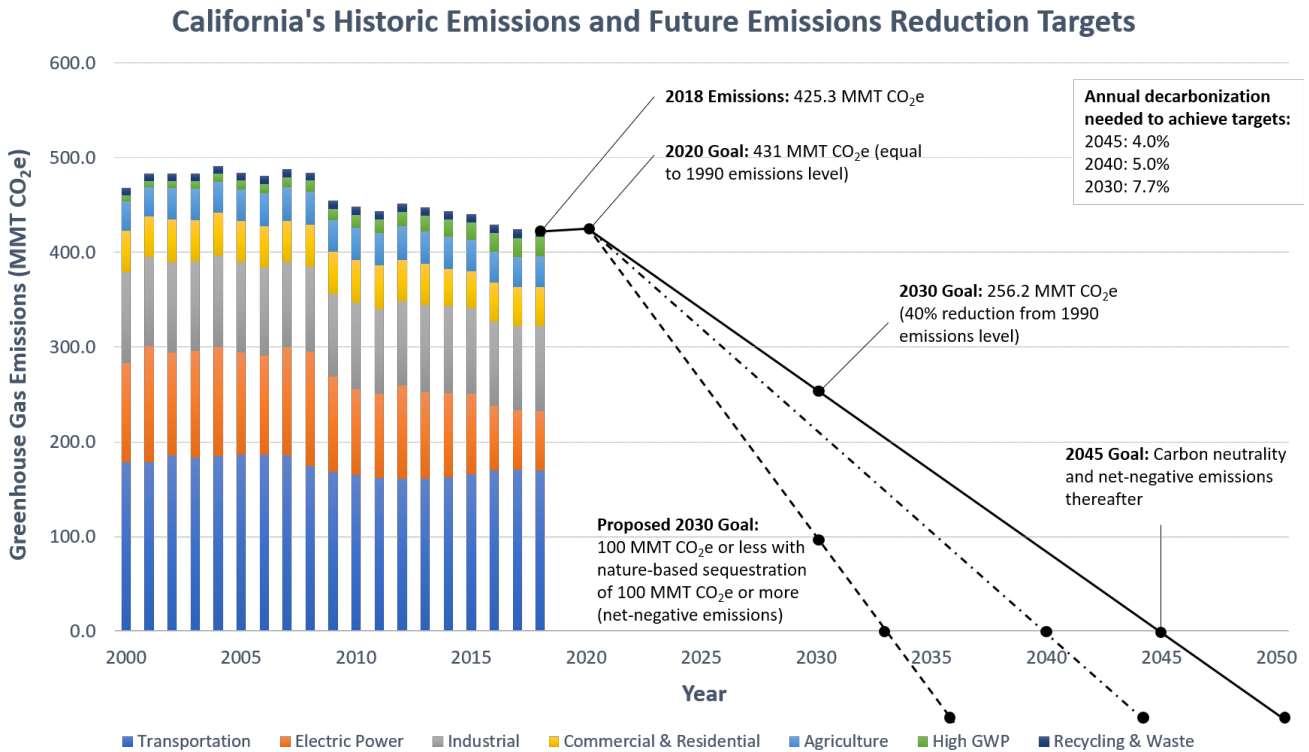

**Figure 1. California's historic emissions and future emission reduction targets. The pathway to carbon neutrality by 2045 (SB100, a 4% annual rate or carbon emissions reductions) is shown (solid line), along with decarbonization pathways that achieve carbon neutrality by 2040 (requiring 5% annual reductions) and 2030 (requiring 7.7% annual reductions) Source: www.theclimatecenter.org**

## 3. California is falling behind

California's once aggressive climate targets are no longer scientifically defensible. California is banking on outdated targets and policies in the face of an accelerating climate crisis, aspiring yet struggling to reach inadequate goals, including 40% below 1990 levels of GHG emissions by 2030, 60% GHG-free electricity by 2030, and carbon neutrality by 2045 (Figure 1).[31] Governor Newsom's recent Executive Order banning the sale of new internal combustion engine cars and light trucks by 2035, although first in the nation, is being outpaced by other governments (the United Kingdom and Norway, for example). Volvo's commitment to producing only Zero Emissions Vehicles (ZEVs) by 2030[32] and Lyft's recent announcement to move to a zero emissions fleet (~2 million drivers and vehicles) by 2030, a deadline that could be accelerated, highlights that faster action is coming.[33]

The nation and other states are advancing decarbonization policies relative to California. For example, President Biden set goals to achieve GHG-free electricity nationally by 2035,[34] and Rhode

---

Island has committed to 100% renewable electricity by 2030, fifteen years ahead of California.[35] This leap-frogging is a good and positive step, but it is time for California to establish a new set of targets backed by feasibility studies that take into account the declining cost for clean energy technologies, a social cost of carbon, and the disproportionate health, housing, transportation, and other costs to frontline and other disadvantaged communities.

The actions of other economies are positive steps and highlight new opportunities for California. Finland has committed to net-zero emissions by 2035 without offsets[36] and Norway is banning the sale of new gas-powered cars by 2025.[37] Prime Minister Boris Johnson of the United Kingdom recently announced an increase in its greenhouse gas (GHG) reduction goals to 68% below 1990 levels by 2030,[38] and called for 100% ZEV car sales by 2030.[39] In December 2020, the European Union committed to a 55% reduction from 1990 GHG emissions by 2030.[40]

## 4. Science and technology-based targets propel action

While incremental steps make achieving a 2030 climate target appear to be impossible, it is clear for health, social justice, economic, environmental, and climate reasons, that California must establish much more aggressive 2030 goals. In doing so, California will be well-positioned to capture a large share of the emerging industry and send important market signals. Because of California's unique relationship with the federal government as well as with China, the state can not only leverage its progress and launch new business in state, but also extend its influence to external markets looking to scale up their efforts.

The past offers many examples of rapid change thought to be impossible by setting establishing bold targets from the transformation of the U.S. economy to fight World War II to the creation of effective COVID vaccines.

Targets offer a high-leverage intervention that can lead to a sustained, transformative clean energy transition.[41] California's climate goals clearly have unified and mobilized state efforts, as they continue to today.[42] Moreover, official targets in one locale can embolden others to follow suit. For example, setting goals to phase out internal combustion engine vehicles has spread rapidly across the

---

world in the last five years,[43] and a number of US states have adopted mid-century climate neutrality targets similar to those in California.

Policy experts recently urged President Biden to adopt lessons from more than a decade of effective climate policy in California. As one said, "the Biden team can pay attention to…the importance of setting really ambitious, economy-wide goals…making clear what we're aiming toward."[44]

## 5. Accelerated climate targets support California's health, social justice, and economic aims

More aggressive climate targets will support California in realizing its health, job-creation, and social justice aims, and will enable it to maintain its stature as a global hub of innovation. In California's inner cities and in air-quality non-attainment areas, the opportunities to directly protect human health and ecosystem function are high, and, as documented by state agencies, surprisingly low-cost.[45]

## 6. Solutions exist today

Solar power, now the cheapest form of electricity globally,[46] has fallen in cost by close to 90% over the past decade, and wind by 70%. Energy storage is now falling in cost as fast as solar and wind energy ever have.[47] Technological and economic progress has been so transformative that the *least cost* new energy technology is now solar and wind power across most of the U.S.

Recent solar projects in Mexico, Dubai, India, and China have all reported solar projects with previously unprecedented clean energy costs of under 1 cent/kWh,[48] driving a clean energy construction boom that is resetting the energy landscape. This pattern is good for working America as well: investments in clean energy generate more than twice as many jobs as fossil fuel investments do.[49] In California, an $80 billion investment in meeting the state's climate goals would create 725,000 jobs.[50]

Examples of policies that accelerate GHG emissions reductions, create jobs, reassert national and global leadership, and foster business innovation include: [51] [52]

- Feebates for the purchase of energy consuming devices – vehicles, appliances, travel, and a host of purchases. These fees levied on higher emitting devices are transferred directly as subsidies to similar-in-class devices that are lower emitting.[53]
- Location-specific feed-in tariffs to reward clean energy generation where the highest credits go to generation (and jobs) in low-income, minority, or otherwise disadvantaged communities. While electricity-generation based emissions are no longer the largest source of emissions in California, this policy will hasten decarbonization and direct investment to the most disadvantaged and historically discriminated against communities.[54]
- Clean energy vehicle credits in both home purchases and leases. To avoid historically racist policies that reward affluent homeowners and ignore renters, a credit for clean vehicle purchases or leases, or mass transit or other sustainable mobility use, can be made available to anyone who signs a mortgage or a lease.
- Meeting the challenge of the Biden Administration to implement and utilize a Social Cost of Carbon for all state-level decision-making.[55]
- Deploy soot-free diesel. Retrofit all diesel vehicles fueled by ultra-low sulfur diesel with diesel particulate filters, on the way to electric vehicles and perhaps hydrogen for trucking.
- Ban food waste from landfills. Fully one-third of food waste is dumped in landfills. It, along with other organic waste, is one of the largest sources of methane, equivalent to the emissions of 37 million cars. The edible part of the food should be used to feed the hungry while the non-edible processed through biodigesters to produce renewable fuels.
- Reduce methane and other climate pollutants produced by cattle and other farm animals through feed additives and manure management including biodigesters that can produce renewable fuel as we shift our diet to depend less on farm-raised animals.
- Produce leak free gas pipes, seal leaking pipes and seal orphan wells, to reduce methane from the production, transmission and distribution of fossil gas.

Based on current clean energy trends, California is well-positioned to achieve new low- and zero-carbon targets. The measures above, in addition to bending the warming curve within 10 years, will also eliminate about 75% of the sources of air pollution in California, especially critical for vulnerable populations.

## 7. Recommendations for California

First, we recommend that the State of California retake its climate leadership by aligning its climate goals with the most recent science, firmly planting its stake in the year 2030. This entails dramatically accelerating its carbon neutrality and net-negative emissions goal from 2045 to 2030.[56] This also means changing its current climate goals both in the percentage and the timing: cutting emissions by

80% instead of 40% below 1990 levels and quickening the pace to achieve the goal by 2030 instead of 2050.[57]

Figure 2 shows a potential pathway that results in net-negative carbon emissions by 2030. It proposes reaching this target via dramatic reductions in all sectors, including a 70% reduction in the transportation sector and a 100% reduction in the electricity sector, supported by a wartime-like mobilization of resources and comprehensive policy support.

This scenario is *simply illustrative*: It demonstrates that even with conventional models it is possible to reach the net-negative target, but it does require truly aggressive targets that have either not been achieved before, or have never been sustained, or both. Clearly, California needs unprecedented action across all sectors. In doing so it may unlock positive tipping points not accounted for in this conventional model and discover new routes as it forges the path we have seen in the synergies between pollution reduction and public health, or between clean electricity standards and electric mobility.[3]

---

[57] http://static1.squarespace.com/static/549885d4e4b0ba0bff5dc695/t/54d7f1e0e4b0f0798cee3010/1423438304744/California+Executive+Order+S-3-05+(June+2005).pdf

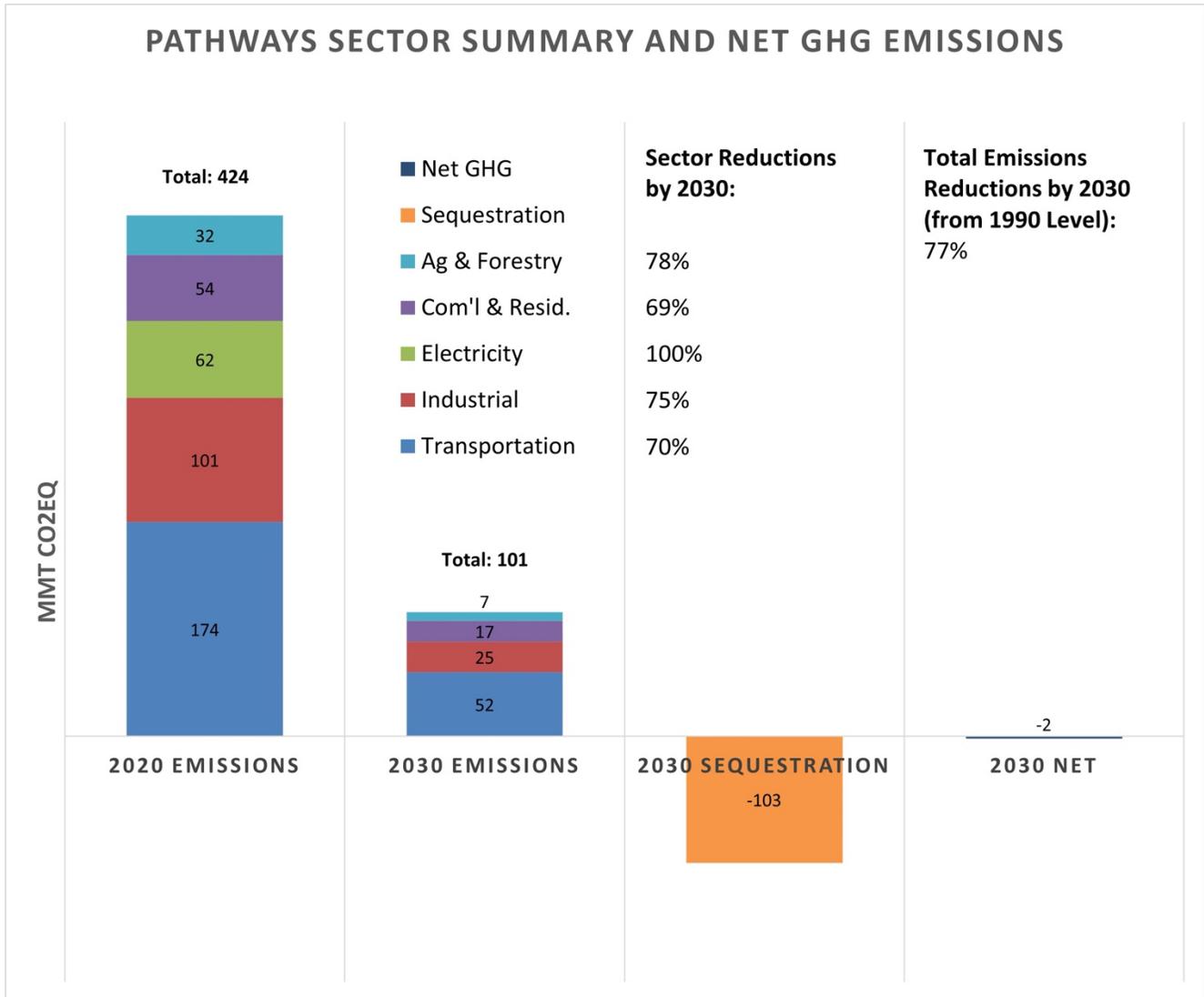

***Figure 2. Pathway of emissions reductions by sector that result in net-negative emissions by 2030. Analysis using The Climate Center GHG Accounting Tool.***

Second, we recommend that the policies enacted to achieve these climate goals conform to three guiding principles:

1. Align with the latest climate science,
2. Ensure a just transition for workers, families, and communities dependent on fossil fuel related industries, and
3. Prioritize investments in lower income communities and communities of color, ensure they are no longer disproportionately harmed by the production and use of fossil fuels, and by climate change impacts, and provide equitable access to climate-friendly solutions.

In making these recommendations, we recognize the enormous political, economic, and technological hurdles to their achievement. However, not overcoming these hurdles will lead to catastrophe.

Surmounting them will save thousands of lives and billions of dollars while securing a healthy, vibrant, and climate-friendly future for all. We must and we can.

## Appendix

Additional findings reported in 2020 make the dire climate forecast increasingly clear:

The ocean's top layer is heating up and thickening due to global warming.[58] Climate change impacts are "proving to be worse that we predicted" and "increased stratification of the ocean could drive a vicious cycle of warming."[59] These changes, which were not expected until later this century, increase threats to fisheries and marine ecosystems as well as the likelihood of more extremes such as severe drought in California.

Recently discovered abrupt permafrost thaw in the Arctic doubles previous estimates of emissions of carbon dioxide ($CO_2$). This thaw also increases the release of methane, which is 86 times more powerful as an agent of warming than $CO_2$ over its ~20-year lifespan in the atmosphere.[60] Themokarst lakes in Alaska were also recently observed draining as the surrounding permafrost walls melted, at rates not expected until the end of the century.[61]

Climate change is causing abrupt changes in dryland regions of the world including western US. These changes are damaging ecosystems where over 2 billion people live and threaten the ability of soils and vegetation in those regions to produce food, sequester carbon, hold water, and sustain biodiversity.[62] Loss of biodiversity and ecosystem function diminish humanity's resilience to growing extremes, reduce food and job security, and increase health risks including from pandemics such as COVID-19.[63]